\documentclass[12pt,a4paper]{article}

  \usepackage{a4wide}
  \usepackage{latexsym}
  \usepackage{epsf}
  \usepackage{amssymb}
  \usepackage{graphicx}
  \usepackage{amsmath, cite}
  \usepackage{amsmath,amssymb,amsthm}
  \usepackage{verbatim}
  \usepackage{hyperref}
\renewcommand{\d}{\textrm{d}}
\newcommand{\e}{\textrm{e}}

\begin{document}
\numberwithin{equation}{section}

\begin{center}

\begin{flushright}
{\small UUITP-19/12\qquad UG-FT-298/12\qquad CAFPE-168/12\qquad
IPhT-t12/050} \normalsize
\end{flushright}
\vspace{0.9 cm}

{\LARGE \bf{Fractional branes, warped compactifications \\\vspace{0.3cm}
and backreacted orientifold planes}}

\vspace{.8 cm} {\large  J. Bl{\aa }b\"ack$^a$, B. Janssen$^b$, T. Van
Riet$^c$, B. Vercnocke$^c$\footnote{{\ttfamily johan.blaback @
physics.uu.se,  bjanssen @ ugr.es, thomas.van-riet, bert.vercnocke @
cea.fr},
}}\\

\vspace{0.75 cm}{$^{a}$ Institutionen f{\"o}r fysik och astronomi,\\
Uppsala Universitet, Box 803,\\   SE-751 08 Uppsala, Sweden}\\

\vspace{0.6 cm}  \vspace{.15 cm} {$^b$ Departamento de F\'{\i}sica
Te\'orica y del Cosmos\\ and Centro Andaluz de F\'{\i}sica de
Part\'{\i}culas Elementales,\\ Universidad de Granada, 18071
Granada, Spain}

\vspace{0.6 cm}  \vspace{.15 cm} {$^c$ Institut de Physique Th\'eorique,\\
CEA Saclay, CNRS URA 2306 \\ F-91191 Gif-sur-Yvette, France}


\vspace{1.2cm}

{\bf Abstract}
\end{center}

\begin{quotation}%
The standard extremal $p$-brane solutions in supergravity are known to
allow for a generalisation which consists of adding a linear
dependence on the worldvolume coordinates to the usual harmonic
function. In this note we demonstrate that remarkably this
generalisation goes through in exactly the same way for $p$-branes
with fluxes added to it that correspond to fractional $p$-branes. We
relate this to warped orientifold compactifications by trading
the D$p$-branes for O$p$-planes that solve the RR tadpole condition. This
allows us to interpret the worldvolume dependence as due to
lower-dimensional scalars that flow along the
massless directions in the no-scale potential. Depending on the details of the fluxes these flows can be supersymmetric domain wall flows. Our solutions provide explicit  examples of backreacted orientifold planes in compactifications with non-constant moduli.

\end{quotation}

\newpage
\tableofcontents
\newpage

\section{Introduction}

Many semi-realistic compactifications of string theory involve brane
or orientifold sources. Such compactifications are necessarily
warped compactifications as the backreaction of the sources
generates (amongst other things) the warp factor. Warping, if strong
enough, can resolve mass hierarchy problems \cite{Giddings:2001yu}
or scale down susy breaking ingredients \cite{Kachru:2003aw} in
order to gain control over corrections to the effective action.
Unfortunately the presence of the warp factor and other backreaction
effects implies we cannot carry out the usual Kaluza--Klein
dimensional reduction to construct the low energy effective theory,
which is also named the \emph{warped effective field theory} (WEFT).
The current understanding of warped effective field theory is
incomplete and the state of the art can be found in the references
\cite{DeWolfe:2002nn, Giddings:2005ff, Kodama:2005cz, Koyama:2006ni,
Frey:2006wv, Douglas:2007tu,  Koerber:2007xk, Douglas:2008jx,
Marchesano:2008rg, Marchesano:2010bs, Martucci:2009sf, Shiu:2008ry,
Frey:2008xw, Douglas:2009zn, Chen:2009zi, Underwood:2010pm,
Grimm:2012rg }.\footnote{ A nice illustration for our incomplete
understanding of warped effective field theory can be found in
\cite{ Underwood:2010pm }.} What is especially worrysome is that one
expects warping corrections to be more relevant when supersymmetry
is broken, such as in the  KKLT proposal for dS vacua
\cite{Kachru:2003aw}. One could speculate that the warping
corrections to the cosmological constant are proportional to the
susy breaking scale and hence the corrections could be equal size as
the bare cosmological constant. The reason warping corrections could
be worse for non-supersymmetric solutions, and especially dS vacua,
are based on experience with existing solutions
\cite{Blaback:2010sj} and on a simple 10-dimensional argument for
classical dS solutions \cite{Douglas:2010rt}. What has been pointed
out in  \cite{Blaback:2010sj} is that for many known
compactifications that are supersymmetric the unwarped limit (in
which sources are smeared) at least captures the correct on-shell
value for the moduli and the cosmological constant. WEFT instead
becomes important when one tries to understand fluctuations around
these supersymmetric (or fake supersymmetric) vacua, for example
when computing moduli masses.  This implies that the warped
effective scalar potential coincides with the unwarped scalar
potential only at those special BPS points in moduli space. We refer
to \cite{Douglas:2007tu} for an explicit example (see figure 1 in
section 4.3 of that paper.)

One method to develop or test WEFT proposals is to construct fully
backreacted ten-dimensional solutions and then dimensionally reduce
them to check the consistency of the lower-dimensional WEFT. This
strategy has been followed in \cite{ Kodama:2005cz,  Kodama:2005fz,
Koyama:2006ni, Frey:2008xw, Blaback:2010sj, Underwood:2010pm}. It is
the aim of this paper to continue with this strategy and take a
first step towards constructing general non-trivial backreacted
solutions that do not describe critical points of the
lower-dimensional theory but dynamical solutions with non-constant
scalars (for instance domain walls and cosmologies). In this paper
we focus on the solutions and their interpretation in the WEFT, but
the actual application to constructing and testing WEFT will be done
elsewhere.

The rest of the paper is organised as follows. In section
\ref{sec:dynamical} we summarise the so called {\it dynamical}
$p$-brane solutions \cite{Kodama:2005cz, Kodama:2005fz,
Koyama:2006ni, Binetruy:2007tu, Underwood:2010pm}. These are
solutions where, in comparison to the standard $p$-brane solutions,
a linear, worldvolume dependent function $H^W$  has been added  to
the harmonic function $H$. We then extend these solutions in section
\ref{sec:genfrac} to also include fluxes, also known as {\it
fractional} branes. Solutions of this type are known ($p=3$ in
\cite{Kodama:2005fz}, and $p=6$ in \cite{Janssen:1999sa}), but we
generalise the solutions to also include $p=1,2,4,5$ in a single
framework. We present the general solution and give some explicit
examples for general $p$. The paper then continues in section
\ref{sec:warped} with a discussion on how these solutions relate to
warped compactifications. The idea is that once fluxes are added one
can trade the D$p$-branes for O$p$-planes and make the internal
space compact. We use this to construct lower-dimensional effective
theories in $p+1$ dimensions, in the smeared limit. We show that the
worldvolume dependent function $H^W$ will imply a running of the
scalars of the effective theory, and also discuss how $H^W$ is
sometimes necessary to assure supersymmetry. The results and
possible implications are then summarised in section
\ref{sec:outlook}. For the reader interested in verifying the
solutions of sections \ref{sec:dynamical} and \ref{sec:genfrac}, we
present the expression for the Ricci tensor in the appendix.

\section{Warm up: dynamical branes}\label{sec:dynamical}

In this section we recall the standard extremal $p$-brane solutions in IIA and IIB for $p=0, \ldots, 6$
and an extension thereof constructed in \cite{Kodama:2005cz,
Kodama:2005fz, Koyama:2006ni, Binetruy:2007tu, Underwood:2010pm},
which has been named ``dynamical $p$-branes'' by some authors. We
present the solutions as magnetic solitons in Einstein frame. The
coordinates along the worldvolume are denoted $x^a$ with $a=1,\ldots,
p+1$ and the coordinates along the transversal space are $y^i$ with
$i=1,\ldots, 9-p$. The solutions are then given by the following
expressions:\footnote{We use the conventions and equations of motion as
presented in \cite{Blaback:2010sj}}
\begin{align}
& \d s^2_{10} = H^{\frac{p-7}{8}} \d s^2_{p+1} +
H^{\frac{p+1}{8}}\d s^2_{9-p}\,,\label{Ansatz1}\\
& \e^{\phi} =  g_s H^{\frac{3-p}{4}}\,, \label{Ansatz2}\\
& F_{8-p} = -g_s^{\frac{3-p}{4}} \star_{9-p}(\partial_i H \d y^i)\,.
\label{Ansatz3}
\end{align}
Where $\star_{9-p}$ is with respect to the transversal space metric
$\d s^2_{9-p}$.\footnote{For $p=3$ it is understood that we have to
add the self-dual piece to the $F_5$ expression in order to make
$F_5$ self-dual ($F_5 = \star_{10} F_5$). From here on we suppress
from writing this extra term for $p=3$ explicitly.} Both the
worldvolume metric and transversal metric are taken to be Ricci
flat:
\begin{align}
& \d s^2_{p+1}= g^W_{ab}\d x^a\d x^b\,,\qquad R_{ab}(g^W) =0\,,\label{Ansatz4}\\
& \d s^2_{9-p}= g^T_{ij}\d y^i\d y^j\,,\qquad R_{ij}(g^T) =0\,.
\label{Ansatz5}
\end{align}
The function $H$ is, by the off-diagonal components of Einsteins equations, forced to have the form of a sum
\begin{equation}
H(x, y) = H^W(x) + H^T(y)\,,
\end{equation}
where $H^T$ is a general harmonic on the transversal space
\begin{equation}
\nabla_i \partial^i H^T =0\,, \label{harmonic}
\end{equation}
and $H^W(x)$ is a worldvolume-dependent harmonic function of a very special kind
\cite{Binetruy:2007tu}
\begin{equation}
\nabla_a \partial_b H^W =0 \label{harmonic2}\,.
\end{equation}
The most well known solutions are those corresponding to string theory D$p$-branes in flat space, for $p=0,\ldots ,6$. Then $g^W$ is the Minkowski
metric, $g^T$ the metric on flat Euclidean space and the harmonic functions are
\begin{equation}
H^W=0\,,\qquad H^T(r) = 1 + \frac{Q}{r^{7-p}}\,,
\end{equation}
where $r$ is the radial coordinate in the transversal space.

For a flat worldvolume the solution for $H^W$ is most easily
written in Cartesian coordinates
\begin{equation}
H^W =
\sum_a c_a x^a\,,
\end{equation}
where $c_a$ are constants. Since $x^0$ is the time-direction we can
for instance make time-dependent brane solutions in this way
\cite{Gibbons:2005rt}.

\section{Generalised fractional brane solutions}\label{sec:genfrac}

We  extend the above known dynamical $p$-brane solutions to
solutions with extra fluxes. These extra fluxes can be interpreted
in some cases as $(p+2)$-branes wrapping (collapsing) 2-cycles in
the transversal space. Such solutions are also known as fractional
branes, famous examples include fractional D3 branes \cite{Klebanov:2000nc
,Klebanov:2000hb} and fractional M2 branes\cite{Cvetic:2000db}; see \cite{Herzog:2000rz} for a
discussion of fractional brane solutions of various dimensionality.
In this section we construct such solutions with non-zero $H^W$. To
our knowledge most of these solutions are new. The $p=6$ solution
with a specific choice of $H^W$ has been constructed before
\cite{Janssen:1999sa} and the $p=3$ solutions have appeared in
\cite{Kodama:2005fz}.

One way to understand the existence of fractional brane solutions is
by properly interpreting the  Bianchi identity for a $F_{8-p}$ RR
field strength:
\begin{equation}\label{Bianchi}
\d F_{8-p} = H_3 \wedge F_{6-p} + \text{delta-source terms}\,.
\end{equation}
We notice that a suitable combination of $F_{6-p}$ and $H_3$ flux
acts as a regularised magnetic source for $F_{8-p}$, in exactly the same way
as a D$p$-brane source smeared over all transversal directions. This suggest that
a suitable flux choice might be mutually BPS with D$p$-brane sources
and could therefore be added to the usual D$p$-solution. This is an
essential ingredient in for instance the Klebanov--Strassler
background \cite{Klebanov:2000hb} and all its related solutions,
where a suitable combination of $F_3$ and $H_3$ is added to D3 brane
backgrounds.

The suitable choice of fluxes that is mutually BPS with D$p$/O$p$
sources is such that \cite{Blaback:2010sj}
\begin{equation}\label{ISD}
F_{6-p} =  g_s^{-\frac{p+1}{4}}\star_{9-p} H_3\,.
\end{equation}
This is the generalization of the imaginary self-duality (ISD) condition
for the complex three-form for $p=3$. For the other fields in the solution we again take  the Ansatz \ \eqref{Ansatz1}--\eqref{Ansatz3}.

In summary  we have\footnote{A notational subtlety arises when
$p=2$. Then $F_4$ and $F_6$ should not be seen as separate forms. To
write the solution more correct one could for instance just use
$F_4$ and add the Hodge dual expression for $F_6$ to it.}
\begin{align}
& \d s^2_{10} = H^{\frac{p-7}{8}} \d s^2_{p+1}(x^a) +
H^{\frac{p+1}{8}}\d s^2_{9-p}(y^i)\,,\label{metric2}\\
& \e^{\phi} =  g_s H^{\frac{3-p}{4}}\,, \\
& F_{8-p} = -g_s^{\frac{3-p}{4}} \star_{9-p}(\partial_i H \d
y^i)\,,\\
& F_{6-p} = g_s^{-\frac{p+1}{4}}\star_{9-p} H_3\,,
\end{align}
where again the transversal and worldvolume metrics are Ricci flat
and the function $H$ is given by a sum $H = H^W(x) +  H^T(y)$. We
still have that  $H^W$ satisfies (\ref{harmonic2}):
\begin{equation}
H^W = \nabla_a \partial_b H^W = 0
\end{equation}
but the equation
for $H^T$ is altered as follows:
\begin{equation}\label{harmonic3}
[\nabla^T]^i\partial_i H^T = -\frac{1}{3!}g_s^{-1}|H_3|_{T}^2,
\end{equation}
up to the source term. The contraction in $|H_3|^2_T$ is done using the transversal
metric $g^T$.

The above Ansatz does not yet solve all equations of motion: the Einstein equations
put non-trivial conditions on the transversal space. It must
have the proper cycles to support the fluxes $F_{6-p}$ and $H$.
Let us construct some explicit simple examples to see how that goes.
As an internal space we choose a Ricci flat cone over a direct
product of Einstein spaces:
\begin{equation}
\d s^2_{9-p} = \d r^2 + r^2 g^{(2)}_{IJ}\d y^{I}\d y^{J} + r^2 g^{(6-p)}_{\alpha\beta}\d y^{\alpha}\d
y^{\beta}\,.\label{eq:DirectProductEinstein}
\end{equation}
For this to be Ricci flat we must have the following curvatures on
the separate Einstein spaces
\begin{equation}
R^{(2)}_{IJ} = (7-p)g^{(2)}_{IJ}\,,\qquad
R^{(6-p)}_{\alpha\beta} = (7-p)g^{(6-p)}_{\alpha\beta}\,.
\end{equation}
One can for instance use $n$-spheres with proper curvature radii.
Note that for $p=5$ this procedure does not work since there will be
a one-dimensional subspace which cannot have non-zero curvature. For
$p=6$ we just have one Einstein space and, if it is taken to be the unit
2-sphere, the transverse space is flat $\mathbb{R}^3$.

We then find that the equations of motion are solved by the natural choice of fluxes
\begin{equation}
F_{6-p} = m\,\epsilon_{6-p}\,,\qquad  H_3 =
(-1)^p g_s^{\frac{p+1}{4}} m  \,r^{p-4} \d r\wedge \epsilon_2\,,
\end{equation}
with $m$ a quantised flux number and $\epsilon_{6-p}$ and $\epsilon_2$ are volume forms on the two Einstein spaces. For $p= 1,2,4,6$, we have explicitly
\begin{equation}\label{KH1}
H^T =  C_2 + \frac{Q}{r^{7-p}} -
\frac{g_s^{\frac{p-1}{2}}\,m^2}{2(p-3)(p-5)}\,\frac 1 {r^{10-2p}}\,,
\end{equation}
where $C_2$ in arbitrary and $Q$ determines the charge of the $p$-brane.
For asymptotically flat backgrounds we can set $C_2=1$. Note that for $p=3$
the last term becomes a logarithm:
\begin{equation}\label{KH2}
H^T =C_2 + \frac{Q}{r^4} + \frac{g_s m^2}{16\, r^4}\Bigl( 4 \,\ln r +
1\Bigr)\,.
\end{equation}

The above solutions with $H^W=0$ described by equations (\ref{KH1},
\ref{KH2}) were discussed before by Herzog and Klebanov in
\cite{Herzog:2000rz}. Note that supersymmetry requires the choice of a suitable
conical internal space. Our simple explicit choice for the
transverse space, a cone over a direct product of Einstein spaces
\eqref{eq:DirectProductEinstein}, breaks all supersymmetry. However, by taking
appropriate internal spaces, it should be possible to construct supersymmetric
solutions with worldvolume dependence $H^W$. We leave the details to future work.
For instance for $p=3$, if we choose the internal space to be a cone over $T^{1,1}$,
there should be a supersymmetric extension of the Klebanov-Tseytlin and its Klebanov-Strassler resolution
with a linear spatial worldvolume dependence.

There is however an exception for the solution with $p=6$. For
$p=6$, reference \cite{Herzog:2000rz} claims that there is no
solution although it was known before by Janssen, Meessen and Ort\'{i}n
(JMO) in a slightly generalised form \cite{Janssen:1999sa}. The
solution in \cite{Janssen:1999sa} was characterised by the following
choice for $H$
\begin{equation}\label{JMO1}
H(r, z) = 1 + \frac{Q}{r} -\frac{1}{6}g_s^{\frac{5}{2}} m^2 r^2 + c z\,,
\end{equation}
where $z$ is a Cartesian spatial coordinate on the D6 worldvolume,
$m$ is the Romans mass and $c$ a constant. We notice that this is
exactly an example of adding a linear worldvolume dependence to the
fractional brane solution.  When $m$
and $c$ both vanish this is the standard extremal D6 solution in IIA
supergravity, which preserves 1/2 of the supersymmetry. When $m\neq
0$ it was shown that maximally 1/4 of the supersymmetry could be
preserved if
\begin{equation}\label{JMO2}
 c=m\,.
\end{equation}
Below we give an interpretation to this worldvolume dependence in
terms of lower dimensional solutions to warped compactifications.
This also allows us to understand when the worldvolume dependence is
required for supersymmetry.

\section{Relation with warped compactifications }\label{sec:warped}

\subsection{General idea}
Ten-dimensional metrics of the kind (\ref{Ansatz1}) can be
regarded as warped compactifications down to $p+1$ dimensions if
the transversal metric, including its conformal factor, can be
interpreted as a metric on a compact space $\mathcal{M}_{9-p}$. For
such compact solutions the integrated Bianchi identity implies a
non-trivial global constraint (Gauss' law):
\begin{equation}\label{tadpole}
\int_{\mathcal{M}_{9-p}} H_3 \wedge F_{6-p} + Q_{\text{total}} =0\,,
\end{equation}
where $Q_{\text{total}}$ is the integral over all delta-sources.

This is the so-called tadpole condition. For fluxes satisfying the
BPS condition (\ref{ISD}) this cannot be satisfied for D$p$-brane
sources. If one instead uses O$p$-plane sources one can satisfy the
tadpole condition and obtain a stable compactification. For $p=3$
this is the well-known GKP compactification \cite{Giddings:2001yu}
and the $p= 1, 2, 4, 5, 6$ solutions are formally related by
T-duality and described in \cite{Blaback:2010sj}.

If one considers these solutions instead in a non-compact setting
and takes them to be spherically symmetric, then these are given by
the brane solutions of the previous section, but with $Q<0$. This
implies that these solutions have unphysical regions, since for
small enough $r$:
\begin{equation}
H^T(r) \to  1 - \frac{|Q|}{r^{7-p}} < 0\,.
\end{equation}
In some occasions one can hope that a lift to M-theory resolves this
unphysical singularity or that the non-zero fluxes in the background
resolve these singularities as in \cite{Saracco:2012wc}. For the
solutions in this paper this does not seem to be the case.

Effectively one can think of a map between two sets of BPS solutions
that one obtains by trading D$p$ sources for O$p$ sources. In practice
this means flipping the sign of the $|Q| r^{p-7}$ term in the
transversal harmonic function $H^T$. For both O$p$ and D$p$ sources, the
fluxes (\ref{ISD}) are mutually BPS with the source. For D$p$-branes
the fluxes and the source gravitationally attract but
electromagnetic forces counter balance this attraction. For O$p$-planes
it is the other way around. In what follows we perform this
flip and look at the fractional brane solutions in which the D$p$
source has been turned into an O$p$ source and the transversal space
is taken to be compact. This allows us to define an effective field
theory on the $(p+1)$-dimensional worldvolume. Due to the presence
of the fluxes these are theories with non-zero scalar potentials for
the moduli.

Clearly, when we consider solutions with $H^W =0$ we just have a
compactification down to $(p+1)$-dimensional Minkowski space. These
are the so-called no-scale Minkowski solutions. What we are after is the interpretation of solutions with $H^W \neq
0$. It is tempting to interpret the $x^a$ dependence as the
dependence of some lower-dimensional scalars on the coordinates
$x^a$. Hence, instead of being stuck at the Minkowski minimum, the
scalars are non-constant. Schematically we have:
\begin{center}
\begin{tabular}{ccl}
$H^W =0$ & $\Longleftrightarrow$& \text{No-scale Minkowski
solution},\\[2mm]
$H^W \neq 0$ & $\Longleftrightarrow$& \text{Non-constant
scalar fields}.
\end{tabular}
\end{center}
We demonstrate this with an explicit example below. We do not expect the general solution with running scalars to lift to the 10D solution with non-zero metric. Only solutions with specific moduli turned on will fit into this class as we explain below.

This interpretation shows that one has to be careful with
interpreting what is the lower-dimensional metric in
(\ref{Ansatz1}). When there is $x^a$-dependence then $\d
s^2_{p+1}(x^a)$ is the Minkowski metric in our solutions. However,
it should not be interpreted as the lower-dimensional metric because
the volume modulus is $x^a$-dependent and one needs to correct for
this if one goes to the Einstein frame metric. Therefore the
lower-dimensional metric is conformal Minkowski, with a conformal
factor related to the volume modulus. This makes sense since a
lower-dimensional Minkowski metric is inconsistent with the
assumption of flowing scalar fields.

Special choices of $H^W$ lead to different lower-dimensional
solutions. For example, when $H^W$ is linear in time, this
corresponds to a specific FLRW solution. When it is linear in a
spatial coordinate instead it describes a domain wall like solution
\begin{align}
&H^W \propto t  \qquad \Longleftrightarrow \qquad \text{FLRW
compactification},\label{FLRW}\\
& H^W \propto z  \qquad \Longleftrightarrow \qquad \text{domain wall
compactification}.\label{domainwall}
\end{align}
When both $t$ and $z$ dependence is present the solution is more
difficult to interpret.

\subsection{Explicit example}
To illustrate the above we consider a very simple truncation of the
effective theory down to three scalars. We do this by taking simple
fluxes consistent with the BPS relation (\ref{ISD}). We furthermore
perform the dimensional reduction assuming the orientifold is
\emph{smeared} and therefore ignore all issues related to warped
effective field theory. The reason is that the fully localised
(warped) solutions are presented above and hence known. What we want
to demonstrate is that the solutions of the lower-dimensional
smeared compactification give exactly the $x^a$-dependence of the
$H^W$ function in the 10-dimensional solution. The effect of the
localisation (and hence full backreaction) of the O$p$-plane is
simply to add the $H^T$ piece to the solution. See
\cite{Blaback:2010sj} for an extensive discussion on smeared versus
localised orientifold solutions. The dynamical fractional brane
solutions provide new examples in which the localised versus smeared
source limit is understood from a 10-dimensional point of view. We
plan to elaborate on this in future work.

\subsubsection{Three-scalar truncation}

There exist three obvious moduli: the dilaton $\phi$, the volume
modulus of the internal $(9-p)$-dimensional space, called $v$ and
the volume of the cycle wrapped by the $F_{6-p}$-flux, called
$\chi$. The metric Ansatz, in 10-dimensional Einstein frame, is
given by
\begin{equation}
\d s^2_{10} = \exp(2\alpha v)\d s_{p+1}^2 + \exp(2\beta
v)\Bigl\{\exp(\gamma \chi)\d s^2_{3} +\exp(\delta \chi) \d
s^2_{6-p}\Bigr\}\,,
\end{equation}
with the numbers $\alpha, \beta, \gamma, \delta$ chosen such that we
end up in lower-dimensional Einstein frame with canonically
normalised fields:
\begin{equation}
\beta =-\tfrac{(p-1)}{9-p}\alpha \,,\quad \alpha^2
=\tfrac{9-p}{16(p-1)}\,,\quad \gamma =
-\tfrac{\delta(6-p)}{3}\,,\quad \delta^2 = \tfrac{6}{(9-p)(6-p)}\,.
\end{equation}
We have made a very strong simplifying assumption: the $H_3$ field
strength fills the $3$-dimensional subspace with metric $\d s^2_3$
and the $F_{6-p}$ flux fills the subspace with metric $\d
s^2_{6-p}$. This is a very simple flux Ansatz, that allows us to
find a consistent truncation down to the three moduli $\phi, v,
\chi$. This consistent truncation is only possible in the smeared limit. In the warped case there might not exist a simple lower-dimensional truncation and even the degrees of freedom are not clear \cite{Underwood:2010pm}.  Hence it remains to be seen whether any reliable information can be obtained from this truncation. As we check below, this truncation is capable of describing the linear dependence in $H^W$, but not more.

The lower-dimensional action is obtained from a direct dimensional reduction of the type II supergravity action with a smeared O$p$ source
\begin{equation}\label{noscalemodel}
S=\int\sqrt{-g}\Bigl(R-\tfrac{1}{2}(\partial\phi)^2
-\tfrac{1}{2}(\partial\chi)^2-\tfrac{1}{2}(\partial v)^2 - V(\phi,
v,\chi)\Bigr)\,.
\end{equation}
The scalar potential $V$ gets contributions from the fluxes and the
negative orientifold plane tension and has the form of an exact
square (due to the tadpole condition). To write it down in a clean
way we perform the following $SO(3)$ field rotation $(\phi, v,
\chi)\rightarrow (x, u, w)$:
\begin{align}
& \phi = -\tfrac{(p-3)}{4}\sqrt{\tfrac{p-1}{2p}}\,\, x + \tfrac{p+1}{8}\,\,u + \tfrac{p+1}{8}\sqrt{\tfrac{3(6-p)}{p}}\,\,w \,,\\
& v = -(p+1)\sqrt{\tfrac{9-p}{32 p}}\,\,x
-\tfrac{p-3}{8}\sqrt{\tfrac{p-1}{9-p}}\,\,u
-\tfrac{p-3}{8}\sqrt{\tfrac{3(6-p)(p-1)}{(9-p)p}}\,\,w\,,\\
& \chi = - \tfrac{1}{2}\sqrt{\tfrac{6(6-p)}{9-p}}\,\,u +
\sqrt{\tfrac{p}{2(9-p)}} \,\,w \,.
\end{align}
The scalar potential then becomes
\begin{equation}\label{potential}
V(x, u, w) = \tfrac{1}{2}\exp(-2\sqrt{\tfrac{p}{2(p-1)}}\,\,x) \,\,
   \Bigr[\mathcal{H}\exp(-u) -
\mathcal{F}\exp(+u)\Bigl]^2 \,,
\end{equation}
where $\mathcal{H}$ and $\mathcal{F}$ represent the flux densities
and are positive numbers. We furthermore observe that the scalar
potential  only depends on two scalars  ($u$ and
$x$)~instead of three. The scalar potential (\ref{potential}) is of
the no-scale type and at the no-scale Minkowski vacuum we have
$ \exp(2u) = \frac{\mathcal{H}}{\mathcal{F}}$,
whereas $x$ and $w$ have arbitrary constant values.

This lower-dimensional action is expected to be a truncation of a half maximal gauged supergravity in $p+1$ dimensions for which the scalar coset is truncated to three scalar fields spanning three flat directions. For the case $p=3$ this half maximal gauged supergravity  has been worked out in all detail in \cite{D'Auria:2003jk}.

For the special case $p=6$, which is an O6 compactification to 7
dimensions, the $\chi$ modulus is absent and we have a two-scalar
truncation instead. From here on we keep the notation general and
include $\chi$ with the understanding that it is absent when $p=6$.
Our flux Ansatz is the most general one  for $p=6$, but for $p<6$
more general flux choices exist. This will be relevant later when we
discuss the supersymmetry of the solutions. It turns out that the
current flux choices do not allow for supersymmetric Minkowski
solutions. However, they allow for domain wall flows that can be
fake and genuine supersymmetric. In the next subsection we construct
some of the domain wall solutions as they will be related to the
fractional brane solutions with non zero $H^W$.

\subsubsection{Some special domain wall flows}

A domain wall Ansatz is given by
\begin{equation}\label{domainwallAnsatz}
\d s^2_D = f^2(y)\d y^2 + g(y)^2\eta_{\mu\nu}\d x^{\mu}\d x^{\nu}\,,
\end{equation}
where $\eta$ is the metric on Minkowski space. We furthermore assume
that the scalars only depend on $y$. Note that $f(y)$ is a gauge
choice that corresponds to redefining the $y$-coordinate. For the presentation
of the solution, we follow Bergshoeff et.~al.~\cite{Bergshoeff:2004nq} and
choose the gauge $f = g^{2-p}$ to present the solutions. When we  uplift to 10
dimensions below, we prefer to choice the conformal gauge $f=g$.
To distinguish between both coordinates we use the coordinate
$y$ in the Bergshoeff gauge and the coordinate $z$ in the conformal
gauge.

In the language of \cite{Bergshoeff:2004nq} our effective action is
a consistent truncation of an $SO(2)$ gauging of maximal supergravity
in 7 dimensions. The real effective theory of the smeared O6
compactification should however be a half-maximal supergravity in 7
dimensions.  What counts here is that the bosonic fields of our truncation
fit into the formalism of \cite{Bergshoeff:2004nq} where the solution is given in
terms of two harmonic functions \cite{Bergshoeff:2004nq}
\begin{equation}
h_1 = 2\mathcal{H} y + \ell^2_1\,,\qquad h_2= 2\mathcal{F} y +
\ell^2_2 \,,
\end{equation}
as follows
\begin{align}
& g = (h_1h_2)^{\frac{1}{4(p-1)}}\,, \label{susy1}\\
& \exp( x)= (h_1h_2)^{\sqrt{\tfrac{p}{8(p-1)}}}\,, \label{susy2}\\
& \exp(u) = (\frac{h_1}{h_2})^{1/2}\,. \label{susy3}
\end{align}
By shifting the coordinate $y$ we can always set one of the two
$\ell_i$ constants equal to zero.

A nice property of the solutions with both $\ell_i^2=0$, and
$\mathcal{H}=\mathcal{F}$,  is that they allow for a simple
non-supersymmetric generalisation. This generalisation consists in
deforming the supersymmetric solution (\ref{susy1}, \ref{susy2},
\ref{susy3}) by rescaling the harmonic functions $h_1, h_2$ as
follows
\begin{equation}\label{fakesusy}
h_{1,2}\rightarrow a h_{1,2}\,.
\end{equation}
To our knowledge these solutions are new.

Let us now uplift the solutions to 10 dimensions. With the above
reduction Ansatz this is straightforward. In what follows we
restrict to $p=6$. As we anticipated, when the sources are smeared
over the internal space ($H^T=0$), the domain wall solution exactly
gives the worldvolume dependence of the localised fractional brane
solutions, after the coordinate redefinition
\begin{equation}
y = \frac{\mathcal{F}}{2}\,z^2\,.
\end{equation}
and
\begin{equation}\label{constants}
c= a {\mathcal{F}}\,.
\end{equation}
As we explained before the coordinate $z$ is the coordinate for
which the domain wall metric is conformal to Minkowski ($f=g$).

Hence we have reproduced the linear space-time dependence $H^W$ of the 10D warped solution using the smeared approximation. The linear dependence is only recovered for the solutions with both $\ell_i^2 =0$ and $\mathcal{H}=\mathcal{F}$. These are the solutions with only one scalar, $x$ , running and all other scalars fixed. More involved solutions will lift to other 10D solutions, which have not yet  been constructed.

\subsubsection{Supersymmetry} Note that the matching condition
(\ref{constants}) is obeyed if one realises that our
notation implied that for $p=6$ the flux density $\mathcal{F}$ is
equal to the Romans mass $\mathcal{F} = m$. Then the supersymmetry
of the domain wall solution corresponds to the supersymmetry
condition for the JMO solution
\begin{equation}\label{SUSYJMO}
a=1 \qquad \Longleftrightarrow \qquad c =m\,.
\end{equation}

\subsection{Interpretation}
The interpretation of dynamical fractional brane solutions as warped
compactifications to $p+1$ dimensions allows a very simple
understanding of:
\begin{enumerate}
\item Why the same linear $H^W$ is still possible
when fluxes are added.
\item When supersymmetry requires the linear dependence in
$H^W$.
\end{enumerate}

Let us start with the first point. Consider dynamical $p$-brane
solutions without fluxes. When interpreted as a warped
compactification\footnote{For the rigorous reader who is worried
about the tadpole condition one can think of placing some
orientifold sources far away from the D$p$ source.}, the absence of
fluxes implies the absence of a tree-level scalar potential; all
scalars are free. Since the $x^a$-dependence for the fractional
branes (the solution with flux) can be interpreted as the running of
the lower-dimensional scalars, the same will be true for the
solutions in which the scalars are free. What is then to be
understood is why both the solutions with a scalar potential and the
solutions without a scalar potential lift to the same 10-dimensional
$H^W$ dependence. Since the $H^W$-dependence is purely generated by
the lift of the scalar fields, it must be that the scalar fields
have the same expression in both the solutions with flux and without
flux. Indeed the domain wall solutions with both $\ell_i^2=0$
(\ref{susy1}, \ref{susy2}, \ref{susy3}) and our non-supersymmetric
generalisation (\ref{fakesusy}) are such that the scalars flow
through the minimum of the scalar potential. More specifically,
during the whole flow we have that
\begin{equation}\label{no-scaleflow}
V = 0 \,,\qquad \partial_u V=0\,,\qquad \partial_x V=0\,.
\end{equation}
This means that \emph{on-shell} there is no difference with the
solutions of the free lower-dimensional theory. We leave the investigation
of more general solutions that do not obey the above condition
(\ref{no-scaleflow}) for future work.

Similarly it is obvious to understand the solutions with the linear
dependence in time (\ref{FLRW}). These are simply the cosmological
solutions that obey (\ref{no-scaleflow}). Since the domain/wall
cosmology correspondence \cite{Skenderis:2006jq} flips the sign of
$V$, but $V=0$ on shell, the same domain wall solutions can be
Wick-rotated to cosmological solutions of the same supergravity
theory. This then explains the linear $t$ behavior. For the case of
$p=3$, which are warped compactifications to $D=4$, these are FLRW
solutions that correspond to ultra-stiff cosmological fluids. In
FLRW cosmological time $\tau$ these have the following scale factor
\begin{equation}
a(\tau)\sim \tau^{\frac{1}{3}}\,.
\end{equation}

Second we discuss supersymmetry. Consider the JMO solution
(\ref{JMO1}) for which we found a match between the original
supersymmetry condition (\ref{JMO2}) discussed from a 10D point of
view \cite{Janssen:1999sa} and the supersymmetry of the
corresponding domain wall solution in the reduced theory
(\ref{SUSYJMO}). Clearly, when the fluxes are zero then
supersymmetry requires $H^W=0$. Hence it is truly the effect of the
fluxes to introduce non-zero $H^W$ in order to satisfy the
supersymmetry rules. Also this can be readily understood from the
point of view of warped flux compactifications. Since the most well-known case is for compactifications down to $D=4$, we treat
the case $p=3$ first.

As we discussed before, the interpretation of the lower-dimensional
Minkowski vacuum is the 10D solution with $H^W=0$. Then it is well
known that the ISD relation for the fluxes (\ref{ISD}) is not
sufficient for supersymmetry. There are extra constraints
\cite{Giddings:2001yu} on the fluxes and on the geometry: the
internal geometry has to be conformal Calabi-Yau and the ISD fluxes
have to point in a specific direction inside the Calabi-Yau space.
In terms of the complex three-form $G = F_3 -ie^{-\phi} H$, one
requires $G$ to be of complexity type $(2,1)$ and primitive. A
general ISD flux allows primitive $(2,1)$ directions and
$(0,3)$-directions. Hence supersymmetry requires vanishing $(0,3)$
fluxes\footnote{The ISD condition here differs with a sign compared
to earlier sections. Compare for example equation (3.6) of
\cite{Blaback:2010sj} (which agrees with the signs used in the
following equation) with (\ref{ISD}).}:
\begin{equation}\label{SUSY1}
\star_6G_3= iG_3\,,\qquad [G_3]_{0,3}=0\,.
\end{equation}
This implies that for ISD fluxes that obey the above relation
(\ref{SUSY1}) we can have 10D-solutions with $H^W=0$ that are
supersymmetric. However, consider what happens when a non-vanishing
$(0,3)$ piece is present. Then supersymmetry is consistent with the
linear dependence in $H^W$. We can see this from the \emph{real}
superpotential $W$
\begin{equation}
W= |\e^{\mathcal{K}}\mathcal{W}|\,,\qquad
\mathcal{W}=\int\Omega\wedge G \,,
\end{equation}
with $\mathcal{K}$ the Kahler potential. A supersymmetric Minkowski solution
satisfies
\begin{equation}
\frac{ \partial W}{\partial \phi^i}=0\,,
\end{equation}
where the index $i$ runs over the scalar fields. When the flux has a
$(0,3)$ piece this relation no longer can be satisfied since $W$ has
no extremum. There is nonetheless another way to obtain
supersymmetric solutions; one can allow the scalar fields to flow
down the superpotential to create supersymmetric domain wall solutions. The supersymmetry
condition then becomes
\begin{align}
& \dot{\phi}^i = - f\, G^{ij}\frac{ \partial W}{\partial \phi^j}\,,\\
& \frac{\dot{g}}{g} = \tfrac{1}{2(p-1)} \,f\,W\,.
\end{align}
where a dot denotes a derivative with respect to the domain wall
coordinate $y$, $G^{ij}$ is the inverse scalar field space metric
and $f, g$ are the metric functions appearing in the domain wall
metric (\ref{domainwallAnsatz}). Therefore we find that, when a
$(0,3)$ piece is present, one necessarily has to allow the scalars
to flow in order to fulfill the supersymmetry conditions. This in
turns implies the lower-dimensional metric is not Minkowski anymore
but instead is of the domain wall type. The reason the
10-dimensional solution seems to have a $(p+1)$-dimensional
Minkowski part (\ref{Ansatz1}) is simply because we used domain wall
coordinates for which the metric is conformal to Minkowski ($f=g$)
and then the conformal part is absorbed in the warp factor (which
indeed depends on all 10-dimensional coordinates).

Let us apply this to the specific choice of fluxes we made in the
explicit example discussed in the previous section. The real
superpotential is
\begin{equation}
W = \exp(-\sqrt{\tfrac{p}{2(p-1)}}x) \Bigr[ \mathcal{H}\exp(-u) +
\mathcal{F}\exp(+u)\Bigl]\,,
\end{equation}
and does not depend on $w$\footnote{ This potential has the typical
no-scale property $V =\tfrac{1}{2}(\partial_u W)^2$, where the
massless fields at the no-scale Minkowski vacuum are $x$ and the decoupled field $w$. Only for $p=3$ this field
corresponds to the volume modulus.}.  Indeed this superpotential does
not allow a supersymmetric extremum since
\begin{equation}
\partial_x W \neq 0
\end{equation}
as both $\mathcal{F}$ and $\mathcal{H}$ are positive. Instead its
supersymmetric domain wall solutions correspond exactly to the
solutions we gave previously, and which reproduced the linear
dependence in $H^W$ of the JMO solution. The question remains as to
what the uplift of the (supersymmetric)-domain wall solutions, for which $V$
and $\partial V$ are non-zero during the flow, correspond to. We
leave this for future investigation.

\section{Outlook}\label{sec:outlook}
Let us conclude this paper by summarizing our results and what we
can learn from it.

We have shown that the generalisation of extremal $p$-brane
solutions to dynamical $p$-branes goes through in exactly the same
way when fluxes are added to it that correspond to fractional
$p$-branes. In practical terms this means that one can add a term
linear in the worldvolume coordinates to the usual ``harmonic
function'' $H$ that defines extremal solutions\footnote{The reason
we put harmonic between quotation marks is that, in presence of
fluxes, the Laplacian of $H$ is non-zero. }. Whenever the extra
fluxes are present the possibility arises to trade the D$p$-branes
for O$p$-planes and trading the non-compact transversal space for a
compact one. This is a simple consequence of the RR tadpole
condition that can be satisfied with O$p$-planes and fractional
brane type of fluxes. This establishes a map between the fractional
brane type solutions in ten dimensions and warped compactifications.
Exactly this map allowed us to interpret why the linear dependence
of the worldvolume coordinates could be added to the harmonic
function. This turns out to correspond to lower-dimensional
solutions that are described by scalar fields that run along the
massless direction in the minimum of the potential, as if the scalar
is effectively free. This map furthermore provides a simple
understanding of the supersymmetry conditions for this
generalisation since it corresponds to the supersymmetry conditions
in warped orientifold compactifications, which were
known.\footnote{This map between extremal $p$-brane solutions and
(warped) flux compactifications fits is similar in spirit to the
recently established link between black hole solutions and flux
compactifications \cite{Bena:2011pi,Bena:2012ub}. } Whenever the
fluxes are such that the no-scale Minkowski vacuum is not
supersymmetric then the solution must necessarily contain the
worldvolume dependence in order to preserve supersymmetry. We have
worked this out in detail for the simplest case of D6-branes (O6
planes) with $F_0$ and $H$ flux.

There are several interesting implications of our work. When our
solutions are interpreted as warped compactifications they describe
orientifold solutions with running moduli such that the orientifold
backreaction has been taken into full account, whereas most solutions in the literature  are vacuum solutions with constant scalar fields (see for
instance \cite{Grana:2006kf, Blaback:2010sj, Saracco:2012wc}). Our solutions with running scalars  are a natural $D$-dimensional extension of the existing solutions in four dimensions, most notably the solutions in \cite{Kodama:2005cz, Marchesano:2008rg} and part of the solutions in \cite{Frey:2008xw}.  The construction of such solutions
is relevant for our understanding of flux compactifications since
most flux compactifications are only understood in the limit where the
sources are fully smeared. Such a limit takes into account the
contribution of the tension to the four-dimensional energy and the
contribution to the RR tadpoles, but nothing more. For
supersymmetric or BPS-like no-scale vacua it has been noticed that
the complex structure moduli are nonetheless unaltered by the full
backreaction \cite{Blaback:2010sj}. This is useful since it implies
that at least the value of the cosmological constant and the
position of the moduli can be trusted in the smeared limit. However
this is probably all that can be trusted. Fluctuations around the
vacuum, that for instance informs one about the moduli masses,
cannot be trusted in the smeared limit and a \emph{warped effective
field theory} is therefore required. Furthermore one can expect that
for non-supersymmetric (non-BPS like) solutions the moduli positions
and possibly the vacuum energy do get altered when the backreaction
is taken into account \footnote{We refer to \cite{Blaback:2011nz,
Burgess:2011rv} for some explicit investigations of backreaction for
genuine non-BPS like solutions.}. Since the solutions in this paper
feature running moduli one could wonder whether the running of the
moduli is at all affected by the backreaction of the orientifold.
For that purpose we have to compare the smeared solution from the
localised one. Generically one  does not expect the smeared effective action 
to contain information about the true warped effective action. However, we have found that the linear space-time dependence can be captured in the smeared approximation. This has a simple interpretation. The warped effective potential coincides with the KK-effective potential (the unwarped approximation) only at the minimum of both potentials.. This is at least what concerns the scalar potentials of the lower-dimensional field theories. Besides the scalar potential one also expects changes in the kinetic terms for the scalars, due to warping (in fact the whole effective field theory should be different, even the degrees of freedom might differ \cite{Underwood:2010pm}). It is not clear whether these 10-dimensional solutions might teach us anything about the kinetic terms of the lower-dimensional effective action.

It would be most interesting to construct solutions which are not linear in the space-time coordinates. The interpretation of such solutions would be that they correspond to scalars flowing outside of the minimum of the scalar potential. It is likely that such solutions contain information about the scalar potential of the WEFT and the kinetic terms for the scalars, outside of the no-scale moduli space. We plan to report on this in a future publication.

\section*{Acknowledgments}
We like to thank Iosif Bena, Mariana Gra\~{n}a, Hagen Triendl, Bret
Underwood and Timm Wrase for useful discussions.  We especially thank Brecht Truijen for spotting typos in v2 of this paper. 
JB is supported by the G{\"o}ran
Gustafsson Foundation. BJ is partially supported by the M.E.C. under
contract FIS2010-17395 and by the Junta de Andaluc\'{\i}a groups
P07-FQM 03048 and FQM-6552. TVR and BV are supported respectively by
the ERC Starting Independent Researcher Grant
259133-ObservableString and 240210--String--QCD--BH.

\appendix
\section{Ricci tensor}
Consider a metric Ansatz of the form
\begin{equation}
\d s^2_{10} = e^{2A(x,y)}\d s_{p+1}^2 + e^{2B(x,y)}\d s^2_{9-p},
\end{equation}
where $x$ is an external coordinate, and $y$ an internal. The Ricci
tensor components for this metric are
\begin{align}
R_{\mu\nu} &= \tilde{R}_{\mu\nu} - \e^{2(A-B)} \tilde{g}_{\mu\nu}\left(((p+1) \partial_i A + (7-p) \partial_i B)\tilde{g}^{ij}\partial_j A + \tilde{\nabla}_{y}^2 A\right)\nonumber\\
&\quad - \tilde{g}_{\mu\nu}\left( ((9-p)\partial_\rho B + (p-1) \partial_\rho A) \tilde{g}^{\rho\lambda} \partial_\lambda A + \tilde{\nabla}_x^2 A \right)\\
&\quad +\left( ((9-p)\partial_\mu B + (p-1)\partial_\mu A ) \partial_\nu A - (9-p)(\partial_\mu B - \partial_\mu A)\partial_\nu B \right)\nonumber\\
&\quad -\left( (9-p)\tilde{\nabla}_\mu \partial_\nu B + (p-1) \tilde{\nabla}_\mu \partial_\nu A \right)\,, \nonumber\\
R_{\mu i} &= -p\, \partial_\mu \partial_i A - (8-p) \partial_\mu \partial_i B + 8\, \partial _\mu B \partial_i A\,,\\
R_{ij} &= \tilde{R}_{ij} - \e^{2(B-A)} \tilde{g}_{ij}\left(((9-p) \partial_\alpha B + (p-1) \partial_\alpha A)\tilde{g}^{\alpha\beta}\partial_\beta B + \tilde{\nabla}_{x}^2 B\right)\nonumber\\
&\quad - \tilde{g}_{ij}\left( ((p+1)\partial_k A + (7-p) \partial_k B) \tilde{g}^{kl} \partial_l B + \tilde{\nabla}_y^2 B \right)\nonumber\\
&\quad +\left( ((p+1)\partial_i A + (7-p)\partial_i B ) \partial_j B - (p+1)(\partial_i A - \partial_i B)\partial_j A \right)\\
&\quad -\left( (p+1)\tilde{\nabla}_i \partial_j A + (7-p) \tilde{\nabla}_i
\partial_j B \right)\,. \nonumber
\end{align}
Both the metric and Ricci tensor are symmetric under the change
\begin{equation}
\left.\begin{array}{c} A(x,y)\\ p+1\\ x^\mu \end{array}\right\}
\leftrightarrow \left\{ \begin{array}{c} B(x,y)\\ 9-p\\ y^i
\end{array} \right.
\end{equation}

\bibliography{refs}

\bibliographystyle{utphysmodb}

\end{document}